\documentclass[prd,11pt,floatfix,superscriptaddress]{revtex4}

\usepackage{latexsym,amsbsy}
 \usepackage[dvips]{graphicx}
 \DeclareGraphicsExtensions{.eps, .jpg}

\begin{document}
\title{\boldmath
How predictions of cosmological models depend on  Hubble parameter data sets}

\author{G. S. Sharov} 
\author{and V. O. Vasiliev}

\affiliation{Tver state university\\ 170002, Sadovyj per. 35, Tver, Russia}

\email{Sharov.GS@tversu.ru} 



\begin{abstract}
 We explore recent estimations of the Hubble parameter $H$ depending on
redshift $z$, which include 31 $H(z)$ data points measured from differential ages of
galaxies and 26 data points, obtained with other methods. We describe these data
together with Union 2.1 observations of Type Ia supernovae and observed parameters of
baryon acoustic oscillations with 2 cosmological models: the standard cold dark matter
model with the $\Lambda$ term ($\Lambda$CDM) and the model with generalized Chaplygin
gas (GCG). For these models with different sets of $H(z)$ data we calculate
two-parameter and one-parameter distributions of $\chi^2$ functions for all observed
effects, estimate optimal values of model parameters and their $1\sigma$ errors. For
both considered models the results appeared to be strongly depending on a choice of
Hubble parameter data sets if we use all 57  $H(z)$ data points or only 31  data points
from differential ages. This strong dependence can be explained in connection with 4
$H(z)$ data points with high redshifts $z>2$.
\end{abstract}

\keywords{cosmological model, Chaplygin gas, Hubble parameter, Type Ia supernovae,
baryon acoustic oscillations}

 \maketitle \flushbottom

\section{Introduction}\label{Intr}

The latest astronomical observations and their astrophysical interpretation
\cite{Planck15} let cosmologists conclude that our Universe demonstrates  accelerated
expansion and it contains $\simeq4$\% of visible baryonic matter, about 26\% of cold
dark matter  and $\simeq70$\% of dark energy (DE). The visible and  dark matter have
properties of cold dust with close to zero pressure. However dark energy has another
equation of state with large negative pressure $p_{DE}$ close to its energy density
$-\rho_{DE}$ with minus sign. Such a form of matter is considered as a source of the
current cosmological acceleration, it helps us to construct a model that can describe
all available now observational data and restrictions
\cite{Planck15,Planck13,WMAP,SNTable}.

The simplest way to modify the Einstein theory of gravitation and to include dark energy
with the mentioned properties is to add the $\Lambda$ term into the Einstein equations.
In this case cosmological solutions can demonstrate accelerated expansion.  The
resulting dynamical equations may be also obtained, if we add the dark energy component
with the equation of state  $p_{DE}=-\rho_{DE}$ to the usual visible matter and cold
dark matter components. This cosmological model is called $\Lambda$CDM (the $\Lambda$
term with cold dark matter), it is now the most popular and usually considered as the
standard model in interpretation of  observational data \cite{Planck15,Planck13,WMAP}.

However, the $\Lambda$CDM model has some problems, in particular, vague nature of dark
energy and dark matter, the fine tuning problem for the small observed value of
$\Lambda$ and the coincidence problem with surprising proximity of DE and matter
contribution in total energy balance nowadays \cite{Clifton,BambaCNO12}. Due to these
reasons cosmologists suggest a lot of alternative models (see reviews
\cite{Clifton,BambaCNO12,NojOdinFR}), in particular, scenarios with nontrivial equations
of state \cite{KamenMP01,BentoBS02,ShV14,Sharov16}, with interaction between dark
components \cite{Bolotin2015,PanBCh2015,SharovBPNCh2017,PanSharov2017}, with $F(R)$
Lagrangian \cite{NojiriO:2007,ElizaldeNOSZ11,OdintsovSGS2017}, additional space
dimensions \cite{GrSh13} and many others.

In particular, in this paper together with the $\Lambda$CDM model we consider the model
with generalized Chaplygin gas (GCG) \cite{KamenMP01,BentoBS02,ShV14,Sharov16}. In this
model two dark  fluids --- dark energy and dark matter are unified and represented as
one dark component (generalized Chaplygin gas) with the following equation of state
connecting energy density $\rho_{g}$ and pressure $p_{g}$:
 \begin{equation}
p_{g}=-B\,\rho_{g}^{-\alpha}. \label{EoSGCG}
  \end{equation}
Here $B$ and $\alpha$ are positive constants. This fluid generalizes the classical
 Chaplygin gas \cite{KamenMP01} with the equation of state $p=\mbox{const}/\rho$.

For the models  $\Lambda$CDM and GCG in this paper we calculate limitations on model
parameters determined from available recent observations including the Type Ia
supernovae data (SN Ia) from Union 2.1 satellite \cite{SNTable}, observable parameters
baryon acoustic oscillations (BAO) 
and we pay
special attention to different data sets of the Hubble parameter estimations $H(z)$.

Type Ia supernovae are usually considered as standard candles in the Universe, because
they give possibility for each event to determine its epoch and the distance (luminosity
distance) to this object. Supernova is an exploding star with huge energy release,
creating a shock wave on the expanding shell \cite{SNKirshner09}. They are observed in
rather far galaxies because of their giant luminosity. All supernovae are classified in
correspondence with time dependence of the their  brightness (the light curve) and their
spectrum. In particular, stars of Type I have hydrogen-deficient optical spectrum and
they belong to Type Ia  subdivision,  if they also have strong absorption near the
silicon line 615 nm. For Type Ia supernovae astronomers can definitely determine their
luminosity distances from light curves. In this paper Sect.~\ref{Observ} we use the
Union 2.1 compilation \cite{SNTable} with 580 SN Ia.

The observable effect of baryon acoustic oscillations (BAO) is generated by acoustic
waves with ions (baryons), which propagated in the relativistic plasma before the
recombination epoch and stopped after  the drag era corresponding to $z_d\simeq 1059.3$
\cite{Planck15}. This effect is observed as disturbances (a bump) in the correlation
function of the galaxy distribution at the sound horizon scale $r_s(z_d)$
\cite{Planck15,Eisen05}. In Sect.~\ref{Observ} we analyze two types of observational
manifestations the BAO effect from  Refs.~\cite{Percival09}\,--\,\cite{Bautista17}, in
particular, estimations of the Hubble parameter $H(z)$ for different redshifts $z$
\cite{ChuangW12}\,--\,\cite{Bautista17}.

The Hubble parameter $H$ is the logarithmic derivative of the scale factor $a$ with
respect to time $t$, redshift $z$ is also expressed via  $a$
\begin{equation}
 H = \frac{\dot{a}}{a},\qquad z=\frac{a_0}a-1=\frac1a-1,
 \label{Hz}
  \end{equation}
 if we choose here and below the value $a$ nowadays: $a_0=a(t_0)=1$.

The Hubble parameter $H(z)$ as the function of $z$ may be estimated with different
methods: in addition to the mentioned BAO effects
\cite{ChuangW12}\,--\,\cite{Bautista17} (26 data points) we also have the $H(z)$ data
measured from differential ages of galaxies \cite{Simon05}\,--\,\cite{Ratsimbazafy17}
(31 data points  are tabulated  Sect.~\ref{Observ}).

In this paper we compare different approaches in choosing  $H(z)$ data, make
calculations with  all 57 $H(z)$ data points or only 31 points from differential ages
and demonstrate for 2 popular cosmological models $\Lambda$CDM and GCG that predictions
of optimal model parameters strongly depend on a considered Hubble parameter data set.

In  Sect.~\ref{Models} we make a brief review of the models $\Lambda$CDM and GCG and
their dynamics, in Sect.~\ref{Observ} describe observational data and in
Sect.~\ref{Results} we demonstrate and analyze the results of  our calculations.

\section{Models}\label{Models}

For the $\Lambda$CDM model and the model with generalized Chaplygin gas (GCG) the
dynamical equations are deduced from the Einstein equations for the Robertson-Walker
metric with the curvature sign $k$
 $$ 
 ds^2 = -dt^2+a^2(t)\Big[(1-k r^2)^{-1}dr^2+r^2
 d\Omega\Big]
 $$  
 and may be reduced to the system
\begin{eqnarray}
3\frac{\dot{a}^2+k}{a^2}=8\pi G\rho+\Lambda,\label{EqFried}\\
 \dot{\rho}=-3\frac{\dot{a}}{a}(\rho+p).\label{Eqcont}
\end{eqnarray}
 Here the dot denotes the time derivative, $\rho$ and $p$ are correspondingly the energy density and pressure of all matter,
$G$ is the Newtonian gravitational constant, the constant $\Lambda$ equals zero for the
GCG model, the speed of light $c=1$.
 Eq.~(\ref{Eqcont}) is the consequence of the continuity
condition $\nabla_\mu T^{\mu}_\nu=0$.

For both considered models we can neglect the fraction of relativistic matter (radiation
and neutrinos), because the radiation-matter ratio is rather small $\rho_r/\rho_m\simeq
3\cdot10^{-4}$ \cite{Planck15} for observable values $z\le2.36$.

In the $\Lambda$CDM model baryons and dark matter may be considered as one component
with density $\rho=\rho_b+\rho_{dm}$ that behaves like dust because of zero pressure
$p=0$. In this case we use the solution
 $\rho/\rho_0=(a/a_0)^{-3}$ of Eq.~(\ref{Eqcont}) and
rewrite the Friedmann equation (\ref{EqFried}) in the form
 \begin{equation}
\frac{H^2}{H_0^2}=\Omega_m a^{-3}+
 \Omega_\Lambda+\Omega_ka^{-2}=\Omega_m (1+z)^3+
 \Omega_\Lambda+\Omega_k(1+z)^2.
  \label{EqLCDM} \end{equation}
  We divided Eq.~(\ref{EqFried}) by $3H_0^2$, used Eq.~(\ref{Hz}) and the following notations
for the present time fractions of matter, dark energy ($\Lambda$ term) and curvature
correspondingly:
 \begin{equation}
\Omega_m
 =\frac{8\pi G\rho(t_0)}{3H_0^2},\qquad
\Omega_\Lambda=\frac{\Lambda}{3H_0^2},\qquad
 \Omega_k=-\frac{k}{H_0^2}.
 \label{Omega1} \end{equation}
 These values are  connected by the equality
 \begin{equation}
\Omega_m+\Omega_\Lambda+ \Omega_k=1,
 \label{sumOm} \end{equation}
resulting from Eq.~(\ref{EqLCDM}) if we fix $t=t_0$. Thus, in description of the
mentioned  observational data the  $\Lambda$CDM  model  has 3 independent parameters:
$H_0$, $\Omega_m$ and $\Omega_\Lambda$ (or $\Omega_k$).

The GCG model includes two matter components: baryons and the generalized Chaplygin gas,
the common density is $\rho=\rho_b+\rho_g$. Unlike the $\Lambda$CDM in the GCG model one
should separately consider baryonic matter (it may include some part of cold dark
matter) and introduce the corresponding fraction
 $$\Omega_b=\frac{8\pi G\rho_b(t_0)}{3H_0^2}$$
 as an additional model parameter.  However in Ref.~\cite{Sharov16} we demonstrated,
that results of calculations very weakly depend on $\Omega_b$. So in this paper we
consider the simplified model with one (gas) component and suppose  $\Omega_b=0$ or
$\rho=\rho_g$. In this case one can substitute the equation of state (\ref{EoSGCG}) into
Eq.~(\ref{Eqcont}), integrate it and obtain the following consequence of  the Friedmann
equation (\ref{EqFried}) \cite{BentoBS02,ShV14,Sharov16}:
\begin{equation}
\frac{H^2}{H_0^2}=
 \Omega_k a^{-2}+
 (1-\Omega_k)\Big[B_s+(1-B_s)\,a^{-3(1+\alpha)}\Big]^{1/(1+\alpha)}.
  \label{EqGCG}\end{equation}
  Here the dimensionless parameter $B_s=B\rho_0^{-1-\alpha}$ is used
instead of $B$. If we exclude the mentioned above parameter $\Omega_b$,  the GCG model
will have 4 independent parameters: $\alpha$, $B_s$, $\Omega_k$ and  $H_0$.

\section{Observational data}\label{Observ}

\subsection{Supernovae Ia data}

In Sect.~\ref{Intr} we briefly mentioned the observational data under investigation and
here we describe details. For Type Ia Supernovae (SN Ia) we use  $N_{SN}=580$ data
points from the table \cite{SNTable} after the  Union 2.1 satellite investigation. This
compilation provides observed (estimated) values of distance moduli  $\mu_i=\mu_i^{obs}$
for redshifts $z_i$ in the interval $0 < z_i \leq 1.41$. We fit free parameters of our
models, when compare $\mu_i^{obs}$ with  theoretical values $\mu^{th}(z_i) $ of the
distance moduli, which are logarithms
 $$\mu_i^{th}=\mu(D_L)=5\log_{10}\big(D_L/10\mbox{pc}\big)$$
 of the luminosity distance \cite{Planck15,Clifton}:
\begin{equation}
 D_L(z)=\frac{c\,(1+z)}{H_0}S_k
 \bigg(H_0\int\limits_0^z\frac{d\tilde z}{H(\tilde
 z)}\bigg), \quad S_k(x)=\left\{\begin{array}{ll} \sinh\big(x\sqrt{\Omega_k}\big)\big/\sqrt{\Omega_k}, &\Omega_k>0,\\
 x, & \Omega_k=0,\\ \sin\big(x\sqrt{|\Omega_k|}\big)\big/\sqrt{|\Omega_k|}, &
 \Omega_k<0.
 \end{array}\right.
   \label{DL} \end{equation}

For a cosmological model with theoretical value $H(z)$ (\ref{EqLCDM}) or (\ref{EqGCG})
depending on model parameters $p_1,p_2,\dots$ we calculate the distance $D_L(z)$ and the
corresponding $\chi^2$ function, that measures  differences between the SN Ia
observational data and predictions of a model:
 \begin{equation}
\chi^2_{SN}(p_1,p_2,\dots)=\min\limits_{H_0} \sum_{i,j=1}^{N_{SN}}
 \Delta\mu_i\big(C_{SN}^{-1}\big)_{ij} \Delta\mu_j,
  \label{chiSN}\end{equation}
 where $\Delta\mu_i=\mu^{th}(z_i,p_1,\dots)-\mu^{obs}_i$, $C_{SN}$ is the $580\times580$
covariance matrix \cite{SNTable}. For the Union 2.1 data \cite{SNTable} the standard
marginalization over the nuisance parameter  $H_0$ is required \cite{Sharov16}, it is
made as the minimum over $H_0$ in the expression  (\ref{chiSN}).

\subsection{BAO data}

For baryon acoustic oscillations (BAO) we take into account the values $d_z(z_i)$
\cite{Eisen05}
 \begin{equation}
 d_z(z)= \frac{r_s(z_d)}{D_V(z)},\qquad
  D_V(z)=\bigg[\frac{cz D_L^2(z)}{(1+z)^2H(z)}\bigg]^{1/3}.
 \label{dz} \end{equation}
 They were extracted  for redshifts (redshift
ranges) $z=z_i$  from a peak in the correlation function of the galaxy distribution at
the comoving sound horizon scale $r_s(z_d)$. The value  $z_d$ corresponds to decoupling
of  photons, for the sound horizon scale $r_s(z_d)$  here we use the following fitting
formula \cite{Sharov16}
 \begin{equation}
 r_s(z_d)=\frac{(r_d\cdot h)_{fid}}h,\qquad
(r_d\cdot h)_{fid}=104.57\mbox{ Mpc},\qquad
 h=\frac{H_0}{100\mbox{ km}/(\mbox{s}\cdot\mbox{Mpc})},
 \label{rsh}\end{equation}
 providing true  $h$ dependence of $r_d$. The value $(r_d\cdot h)_{fid}=104.57\pm1.44$ Mpc
is the best fit for the $\Lambda$CDM model \cite{Sharov16}.

In our calculations  we use $N_{BAO}=26$  BAO data points for $d_z(z)$ (\ref{dz}) from
Refs.~\cite{Percival09}\,--\,\cite{Wang17}, tabulated here in Table~\ref{TBAO}. We add 9
new points from Ref.~\cite{Wang17} to 17 ones, which were used earlier in
Refs.~\cite{ShV14,Sharov16,SharovBPNCh2017,PanSharov2017,OdintsovSGS2017}.
 We use the covariance matrix $C_{d}$  for correlated data
from Refs.~\cite{Percival09,BlakeBAO11} described in detail in Ref.~\cite{Sharov16}. So
the $\chi^2$ function for the value (\ref{dz}) yields
 \begin{equation}
 \chi^2_{BAO}(p_1,p_2,\dots)=\Delta d\cdot C_d^{-1}(\Delta d)^T,\qquad\Delta
 d_i=d_z^{obs}(z_i)-d_z^{th}(z_i).
  \label{chiB} \end{equation}

\begin{table}[th]
\centering \caption{Values $d_z(z)=r_s(z_d)/D_V(z)$  (\ref{dz}) with errors and
references} \label{TBAO}
 {\begin{tabular}{||l|l|l|l||l|l|l|l||}
\hline
 $z$  &$d_z(z)$&$\sigma_d$ & Refs             & $z$ &$d_z(z)$&$\sigma_d$  & Refs \\ \hline
 0.106& 0.336  & 0.015 & \cite{Beutler11}     &0.44 & 0.0916 & 0.0071 & \cite{BlakeBAO11}\\ \hline
 0.15 & 0.2232 & 0.0084& \cite{Ross14}        &0.44 & 0.0874 & 0.0010 & \cite{Wang17}   \\ \hline
 0.20 & 0.1905 & 0.0061& \cite{Percival09}    &0.48 & 0.0816 & 0.0009 & \cite{Wang17}   \\ \hline
 0.275& 0.1390 & 0.0037& \cite{Percival09}    &0.52 & 0.0786 & 0.0009 & \cite{Wang17}   \\ \hline
 0.278& 0.1394 & 0.0049& \cite{Kazin09}       &0.56 & 0.0741 & 0.0008 & \cite{Wang17}   \\ \hline
 0.31 & 0.1222 & 0.0021& \cite{Wang17}        &0.57 & 0.0739 & 0.0043 & \cite{Chuang13} \\ \hline
 0.314& 0.1239 & 0.0033& \cite{BlakeBAO11}    &0.57 & 0.0726 & 0.0014 & \cite{Anderson14}\\ \hline
 0.32 & 0.1181 & 0.0026& \cite{Anderson14}    &0.59 & 0.0711 & 0.0010 & \cite{Wang17}    \\ \hline
 0.35 & 0.1097 & 0.0036& \cite{Percival09}    &0.60 & 0.0726 & 0.0034 & \cite{BlakeBAO11}\\ \hline
 0.35 & 0.1126 & 0.0022& \cite{Padmanabhan12} &0.64 & 0.0675 & 0.0011 & \cite{Wang17}    \\ \hline
 0.35 & 0.1161 & 0.0146& \cite{ChuangW12}     &0.73 & 0.0592 & 0.0032 & \cite{BlakeBAO11}\\ \hline
 0.36 & 0.1053 & 0.0018& \cite{Wang17}        &2.34 & 0.0320 & 0.0021 & \cite{Delubac14} \\ \hline
 0.40 & 0.0949 & 0.0014& \cite{Wang17}        &2.36 & 0.0329 & 0.0017 & \cite{Font-Ribera13}\\ \hline
\end{tabular}}
\end{table}

Unlike Refs.~\cite{Sharov16,SharovBPNCh2017,PanSharov2017,OdintsovSGS2017}  we do not
use in this paper the observational value  \cite{Eisen05}
 $$
  A(z) = \frac{H_0\sqrt{\Omega_m}}{cz}D_V(z),
 $$
 because it essentially depends on  $\Omega_m$, however $\Omega_m$ is not the model parameter for the GCG
 model  (see Table~\ref{Estim}).

 \subsection{$H(z)$ data}

The Hubble parameter values $H$ at certain redshifts $z$ can be measured with
two methods: (1) extraction $H(z)$ from line-of-sight BAO data
\cite{ChuangW12}\,--\,\cite{Bautista17} including analysis of correlation functions of
luminous red galaxies \cite{ChuangW12,Oka13}, and (2) $H(z)$ estimations from
differential ages $\Delta t$ of galaxies (DA method)
\cite{Simon05}\,--\,\cite{Ratsimbazafy17}
via Eq.~(\ref{Hz}) and the following relation:
 $$ 
 H (z)= \frac{\dot{a}}{a}= -\frac{1}{1+z}
\frac{dz}{dt} \simeq -\frac{1}{1+z} \frac{\Delta z}{\Delta t}.
 $$ 

The maximal set with  $N_H=57$ recent estimations of $H(z)$ is shown in Fig.~\ref{F1}
and in Table~\ref{TH} below, it includes 31 data points measured with DA method (the
left side) and 26 data points (the right side), obtained with BAO and other methods. The
$\chi^2$ function for the $H(z)$ data is
\begin{equation}
 \chi^2_H(p_1,p_2,\dots)=\sum_{i=1}^{N_H}
 \frac{\big[H_i-H^{th}(z_i,p_1,p_2,\dots)\big]^2}{\sigma_{H,i}^2}.
  \label{chiH} \end{equation}

In papers \cite{SharovBPNCh2017,OdintsovSGS2017} we used only $N_H=30$  $H(z)$ data
points estimated from DA method to avoid additional correlation with the BAO data from
Table~\ref{TBAO}. This consideration should be taken into account in the present paper:
in the next section we calculate separately the $\chi^2$ function with  $N_H=31$ DA data
points from the left column of Table~\ref{TH} ($30$ points from
Refs.~\cite{SharovBPNCh2017,OdintsovSGS2017} and the recent point from
Ref.~\cite{Ratsimbazafy17}) and compare these results with the full $H(z)$ data from
Table~\ref{TH} with  $N_H=57$ data points.

In Fig.~\ref{F1} the  $H(z)$ data points from Table~\ref{TH} estimated with DA and BAO
methods are shown as correspondingly red stars and cyan diamonds. The lines demonstrate
the best fitted $H(z)$ dependence with the optimal parameters  from Table \ref{Estim}
for the $\Lambda$CDM and GCG models with  57 and 31
$H(z)$ data points. 

\begin{figure}[t]
  \centerline{\includegraphics[scale=0.71,trim=12mm 0mm 5mm -3mm]{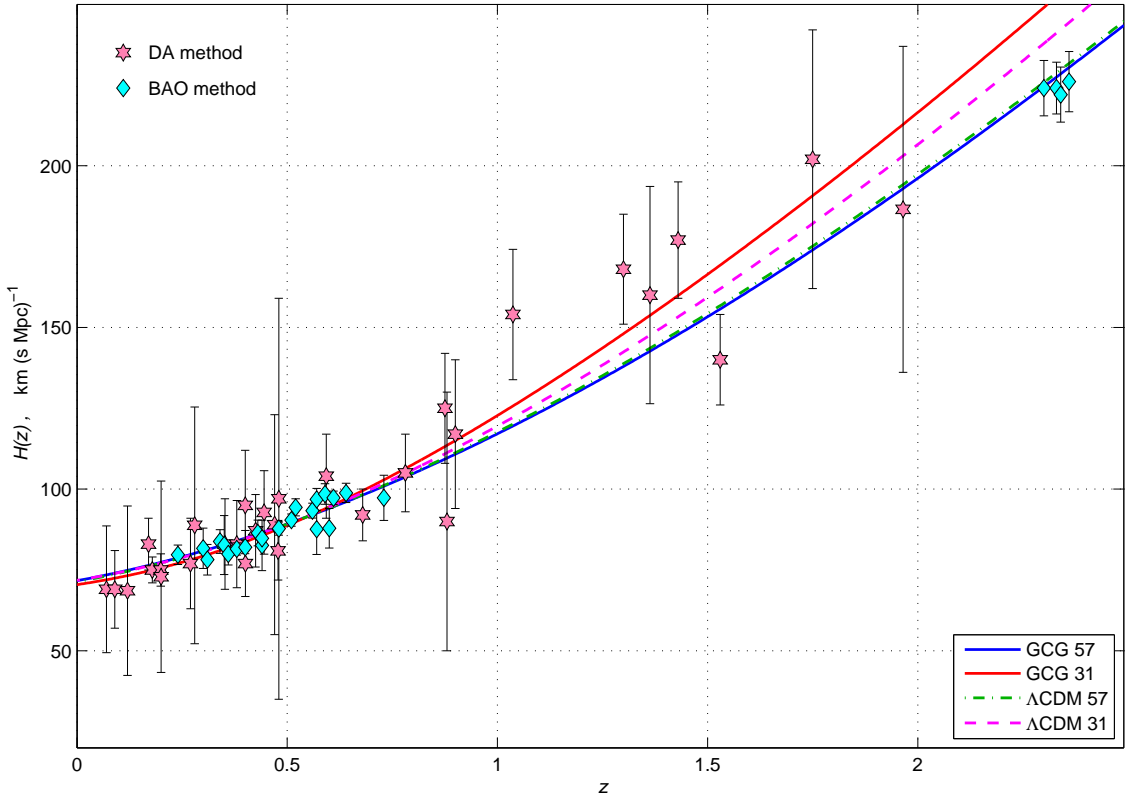}}
\caption{ $H(z)$ data from Table \ref{TH}, stars and diamonds denote data points
correspondingly from DA and BAO methods. The lines are the best fitted for the $\Lambda$CDM and GCG models
with 57 and 31 $H(z)$ data points. }
  \label{F1}
\end{figure}

\section{Results of analysis}\label{Results}

For any cosmological model we investigate the space of its model parameters
$p_1,p_2,\dots$ (they are  $\Omega_m$, $\Omega_\Lambda$,  $H_0$ for the $\Lambda$CDM and
$\alpha$, $B_s$, $\Omega_k$,  $H_0$ for the GCG model) and search the optimal values of
these  parameters, which yield the most successful description of the observational data
from Sect.~\ref{Observ}. To achieve this purpose, for any set of  parameters
$p_1,p_2,\dots$ we use the dependence $H(z)$ (\ref{EqLCDM}) or (\ref{EqGCG}), calculate
the integral in Eq.~(\ref{DL}), the distances $D_L=D_L^{th}(z)$  and $D_V^{th}(z)$
(\ref{dz}), the values  $\mu^{th}$, $d_z^{th}$, the $ \chi^2$ functions $ \chi^2_{SN}$
(\ref{chiSN}), $ \chi^2_{BAO}$ (\ref{chiB}), $ \chi^2_H$ (\ref{chiH}) and the summarized
function
  \begin{equation}
\chi^2_{tot}=\chi^2_{SN}+\chi^2_{BAO}+\chi^2_H.
 \label{chitot} \end{equation}

We search minima of the functions $\chi^2_H$ and $\chi^2_{tot}$ in the parameter spaces
of a model in the two mentioned variants of the $H(z)$ data sets: with all $N_H=57$ data
points from Table~\ref{TH} and with only $N_H=31$ data points from
Refs.~\cite{Simon05}\,--\,\cite{Ratsimbazafy17}, estimated via the DA method.

{  \small
\begin{table}[h]
\caption{Hubble parameter values $H(z)$ with errors $\sigma_H$ from DA and BAO methods.}
 \begin{center}
\begin{tabular}{||l|l|l|c||l|l|l|c||}  \hline
\multicolumn{4}{||c||}{\text{DA method} } & \multicolumn{4}{|c||}{\text{BAO method}}\\
\hline
 $z$  & $H(z)$ &$\sigma_H$  & Refs   &    $z$ & $H(z)$ & $\sigma_H$  & Refs\\ \hline
 0.070 & 69  & 19.6& \cite{Zhang12}  &   0.24 & 79.69& 2.99& \cite{Gazta09}   \\ \hline
 0.090 & 69  & 12  & \cite{Simon05}  &   0.30 & 81.7 & 6.22& \cite{Oka13}     \\ \hline
 0.120 & 68.6& 26.2& \cite{Zhang12}  &   0.31 & 78.18& 4.74& \cite{Wang17}    \\ \hline
 0.170 & 83  & 8   & \cite{Simon05}  &   0.34 & 83.8 & 3.66& \cite{Gazta09}   \\ \hline
 0.1791& 75  & 4   & \cite{Moresco12}&   0.35 & 82.7 & 9.1 & \cite{ChuangW12} \\ \hline
 0.1993& 75  & 5   & \cite{Moresco12}&   0.36 & 79.94& 3.38& \cite{Wang17}    \\ \hline
 0.200 & 72.9& 29.6& \cite{Zhang12}  &   0.38 & 81.5 & 1.9 & \cite{Alam16}    \\ \hline
 0.270 & 77  & 14  & \cite{Simon05}  &   0.40 & 82.04& 2.03& \cite{Wang17}    \\ \hline
 0.280 & 88.8& 36.6& \cite{Zhang12}  &   0.43 & 86.45& 3.97& \cite{Gazta09}   \\ \hline
 0.3519& 83  & 14  & \cite{Moresco12}&   0.44 & 82.6 & 7.8 & \cite{Blake12}   \\ \hline
 0.3802& 83  & 13.5& \cite{Moresco16}&   0.44 & 84.81& 1.83& \cite{Wang17}    \\ \hline
 0.400 & 95  & 17  & \cite{Simon05}  &   0.48 & 87.79& 2.03& \cite{Wang17}    \\ \hline
 0.4004& 77  & 10.2& \cite{Moresco16}&   0.51 & 90.4 & 1.9 & \cite{Alam16}    \\ \hline
 0.4247& 87.1& 11.2& \cite{Moresco16}&   0.52 & 94.35& 2.64& \cite{Wang17}    \\ \hline
 0.4497& 92.8& 12.9& \cite{Moresco16}&   0.56 & 93.34& 2.3 & \cite{Wang17}    \\ \hline
 0.470 & 89  & 34&\cite{Ratsimbazafy17}& 0.57 & 87.6 & 7.8 & \cite{Chuang13}  \\ \hline
 0.4783& 80.9& 9   & \cite{Moresco16}&   0.57 & 96.8 & 3.4 & \cite{Anderson14}\\ \hline
 0.480 & 97  & 62  & \cite{Stern10}  &   0.59 & 98.48& 3.18& \cite{Wang17}    \\ \hline
 0.593 & 104 & 13  & \cite{Moresco12}&   0.60 & 87.9 & 6.1 & \cite{Blake12}   \\ \hline
 0.6797& 92  & 8   & \cite{Moresco12}&   0.61 & 97.3 & 2.1 & \cite{Alam16}    \\ \hline
 0.7812& 105 & 12  & \cite{Moresco12}&   0.64 & 98.82& 2.98& \cite{Wang17}    \\ \hline
 0.8754& 125 & 17  & \cite{Moresco12}&   0.73 & 97.3 & 7.0 & \cite{Blake12}   \\ \hline
 0.880 & 90  & 40  & \cite{Stern10}  &   2.30 & 224  & 8.6 & \cite{Busca12}   \\ \hline
 0.900 & 117 & 23  & \cite{Simon05}  &   2.33 & 224  & 8   & \cite{Bautista17}\\ \hline
 1.037 & 154 & 20  & \cite{Moresco12}&   2.34 & 222  & 8.5 & \cite{Delubac14} \\ \hline
 1.300 & 168 & 17  & \cite{Simon05}  &   2.36 & 226  & 9.3 & \cite{Font-Ribera13}   \\ \hline
 1.363 & 160 & 33.6& \cite{Moresco15}& & & & \\ \hline
 1.430 & 177 & 18  & \cite{Simon05}  & & & & \\ \hline
 1.530 & 140 & 14  & \cite{Simon05}  & & & & \\ \hline
 1.750 & 202 & 40  & \cite{Simon05}  & & & &  \\ \hline
 1.965 &186.5& 50.4& \cite{Moresco15}& & & & \\ \hline
 \end{tabular}
\end{center}
 \label{TH}
 \end{table}
 }

For both considered models we calculate two-parameter distributions of
$\min\chi^2_{tot}$ in planes of two model parameters, for example,
  \begin{equation}
m^\chi_{tot}(p_1,p_2)=\min\limits_{p_3,\dots}\chi^2_{tot}(p_1,p_2,p_3,\dots).
 \label{2param} \end{equation}
 We use this functions to  determine one-parameter distributions and the corresponding likelihood functions:
   \begin{equation}
m^\chi_{tot}(p_j)=\min\limits_{\mbox{\scriptsize other
}p_k}\chi^2_{tot}(p_1,\dots),\qquad {\cal L}_{tot}(p_j)= \exp\bigg[-
\frac{m^\chi_{tot}(p_j)-m_{abs}}2\bigg].
 \label{likeli} \end{equation}
Here $m_{abs}$ is the absolute minimum of $\chi^2_{tot}$.

\begin{figure}[th]
  \centerline{\includegraphics[scale=0.71,trim=8mm 2mm 5mm 4mm]{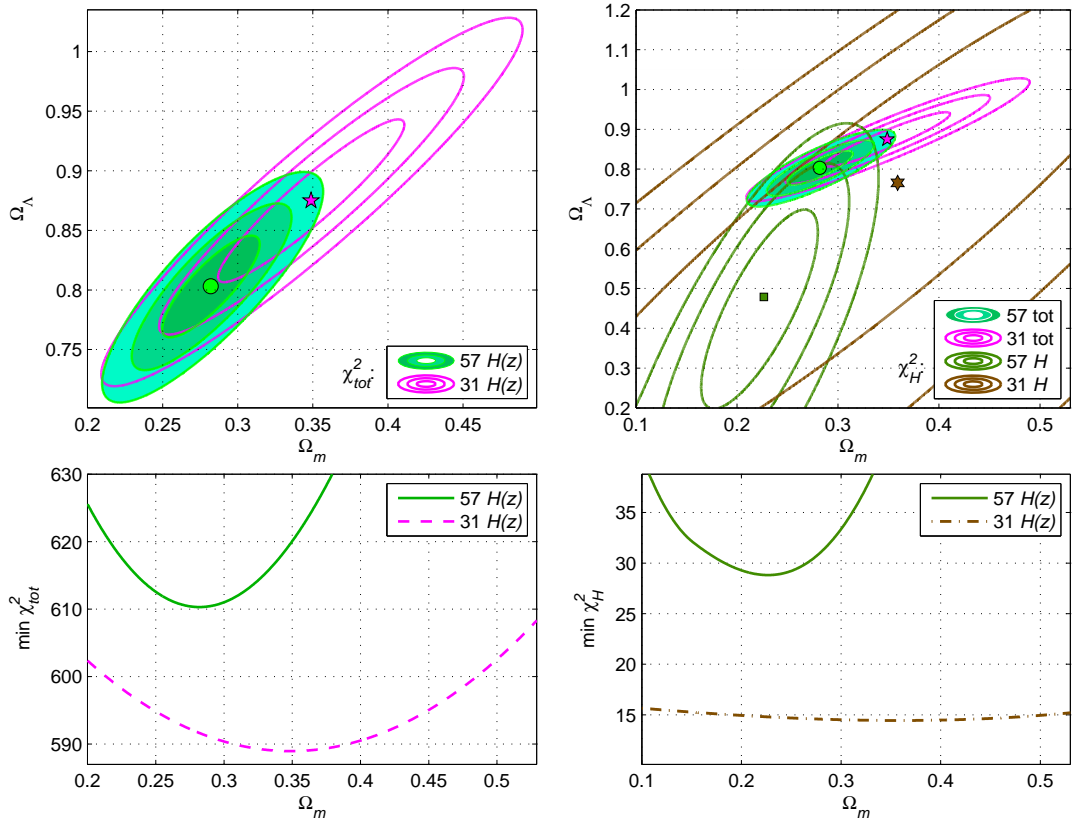}}
\caption{  The $\Lambda$CDM model: $1\sigma$, $2\sigma$ and $3\sigma$ contour plots for
two-parameter distributions $m^\chi_{tot}(\Omega_m,\Omega_\Lambda)$ are drawn in
$(\Omega_m,\Omega_\Lambda)$ plane for 57 and 31 $H(z)$ data points 
 in comparison with  contours for $\min\limits_{H_0}\chi^2_H$ (the
top-right panel). The corresponding one-parameter distributions
$m^\chi_{tot}(\Omega_m)$ and $m^\chi_H(\Omega_m)$  are in the bottom panels.
 }
  \label{F2}
\end{figure}

\begin{figure}[th]
  \centerline{\includegraphics[scale=0.71,trim=6mm 4mm 5mm 11mm]{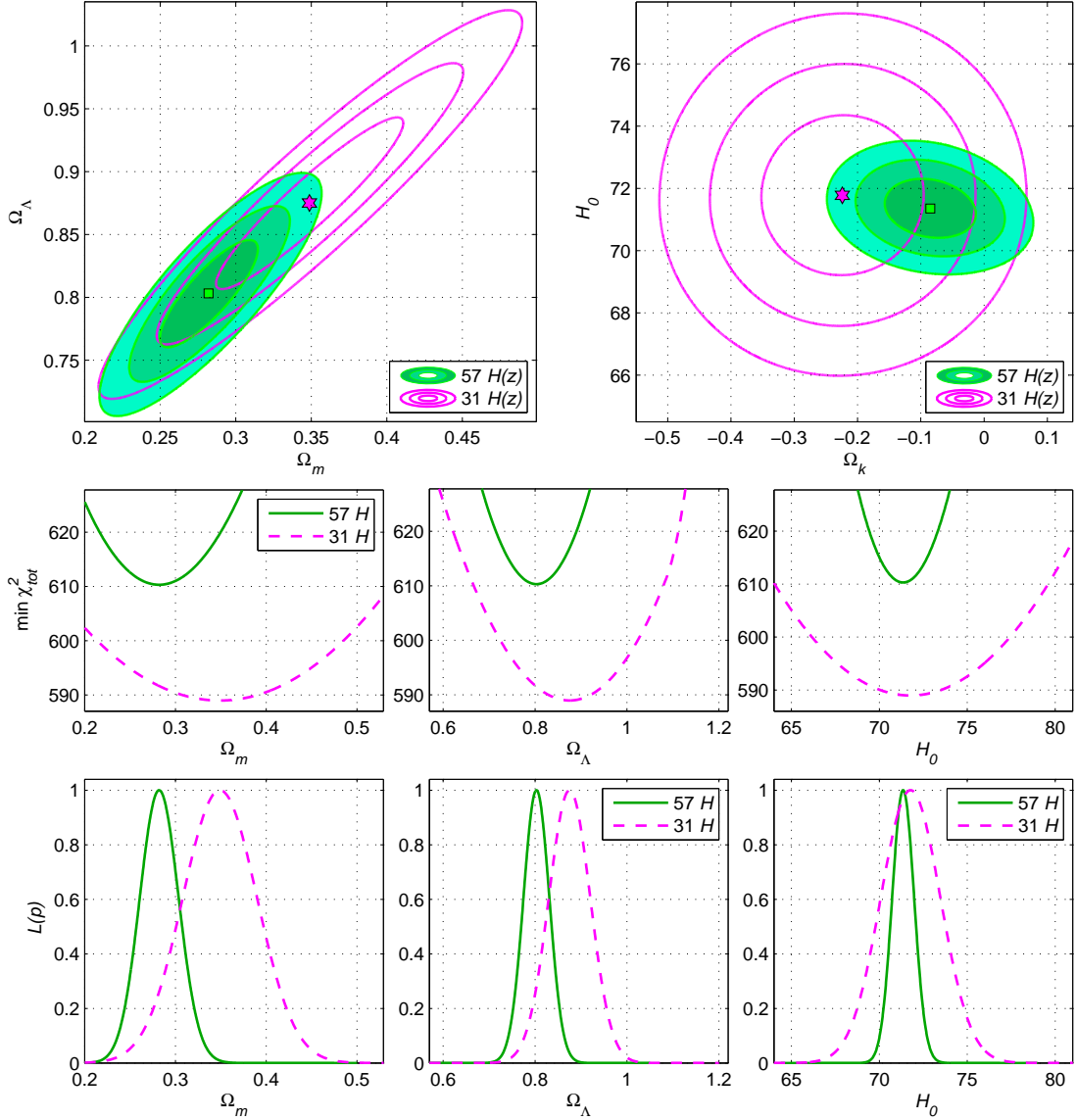}}
\caption{  The $\Lambda$CDM model with 57 and 31 $H(z)$ data points: contour plots in 2
planes, one-parameter distributions and likelihood functions. }
  \label{F3}
\end{figure}

The results of these calculations for the $\Lambda$CDM model with three independent
parameters $\Omega_m$, $\Omega_\Lambda$ and $H_0$ are presented in Figs.~\ref{F2},
\ref{F3} and in  Table~\ref{Estim}. In the  top-left panel of Fig.~\ref{F2} we draw the
contour plots at $1\sigma$ (68.27\%), $2\sigma$ (95.45\%) and $3\sigma$ (99.73\%)
confidence level for the two-parameter distributions (\ref{2param}) of $\chi^2_{tot}$ in
the ($\Omega_m,\Omega_\Lambda$) plane. The green filled contours describe the
$m^\chi_{tot}(\Omega_m,\Omega_\Lambda)$ function for all 57 $H(z)$ data points, the
magenta contours present the case with 31 DA $H(z)$ data points.  Here the function
(\ref{2param}) is
  \begin{equation}
m^\chi_{tot}(\Omega_m,\Omega_\Lambda)=\min\limits_{H_0}\chi^2_{tot}(\Omega_m,\Omega_\Lambda,H_0).
 \label{2paramL} \end{equation}

In the top-right panel of Fig.~\ref{F2} we compare the mentioned  contours for
$\chi^2_{tot}$ (with the same colors) and the similar contours for the function
$\chi^2_H$ (\ref{chiH}), more correctly,
 $$ 
m^\chi_H(\Omega_m,\Omega_\Lambda)=\min\limits_{H_0}\chi^2_H(\Omega_m,\Omega_\Lambda,H_0).
 $$ 
This distribution  includes only $H(z)$ data.

The green circles and magenta stars in Fig.~\ref{F2} denote the minimum points of
$m^\chi_{tot}(\Omega_m,\Omega_\Lambda)$ (and, naturally, for  $\chi^2_{tot}$)
correspondingly for 57 and 31 $H(z)$ data points. Their coordinates (the optimal values
of parameters) are tabulated in Table~\ref{Estim}. In the same way, the minimum points
for $\chi^2_H$ are shown in the top-right panel as the deep green square and brown
hexagram.

In the bottom panels of Fig.~\ref{F2} we  compare the  one-parameter distributions
(\ref{likeli}) $m^\chi_{tot}(\Omega_m)$ and
$m^\chi_H(\Omega_m)=\min\limits_{\Omega_\Lambda}m^\chi_H(\Omega_m,\Omega_\Lambda)$.
These distributions and the corresponding likelihood functions (\ref{likeli}) determine
$1\sigma$ estimates in Table~\ref{Estim} (for $\chi^2_{tot}$).

In Fig.~\ref{F2} we see the interesting phenomenon: the optimal values of parameters
$\Omega_m$, $\Omega_\Lambda$ (and positions of minimum points for $\chi^2$) are
essentially different for the two considered cases with  57 and 31 $H(z)$ data points.
This divergence takes place for $\chi^2_{tot}$ (the left panels in Fig.~\ref{F2}), for
example, these estimations for $\Omega_m$ are correspondingly $\Omega_m=0.282\pm0.021$
and  $\Omega_m=0.349\pm0.041$ (see Table~\ref{Estim}): the last value $0.349$ is beyond
$2\sigma$ confidence level for the $N_H=57$ case. However for $\chi^2_H$ this divergence
is stronger, the correspondent estimations are $\Omega_m=0.227_{-0.041}^{+0.036}$ (for
$N_H=57$) and $\Omega_m=0.359_{-0.232}^{+0.197}$ (for $N_H=31$). This is natural,
because the summands $\chi^2_{SN}+\chi^2_{BAO}$ in $\chi^2_{tot}$ moderate this effect.

\begin{table}[hb]
\caption{Optimal values and $1\sigma$ estimates of model parameters}
\begin{tabular}{||c||c|c||c|c|c||}  \hline
  Model  & $\min\chi^2_{tot}$ & AIC &$H_0$ & $\Omega_k$ & other parameters \\ \hline
 $\Lambda$CDM & 610.31&  616.31& $71.35_{-0.62}^{+0.63}\rule{0mm}{1.2em}$ &$-0.085\pm0.048$ &
 $\Omega_m=0.282\pm0.021$,  \\
 57 $H(z)$  &  &  &  &  &  $\Omega_\Lambda =0.803\pm0.028 $   \\
  \hline
   $\Lambda$CDM & 588.96& 594.96& $71.77_{-1.69}^{+1.70}\rule{0mm}{1.2em}$ &$-0.224_{-0.084}^{+0.085}$ &
 $\Omega_m=0.349\pm0.041$,  \\
 31 $H(z)$  &  &  &  &   &  $\Omega_\Lambda =0.875\pm0.045 $    \\
  \hline
 GCG & 609.94&  617.94& $71.68_{-0.83}^{+0.82}\rule{0mm}{1.2em}$ &$-0.192_{-0.170}^{+0.188}$ &
 $\alpha=-0.124_{-0.138}^{+0.235}$,  \\
 57 $H(z)$  &  &  &  &  &  $B_s =0.705_{-0.044}^{+0.065} \rule{0mm}{1.2em}$   \\
  \hline
   GCG & 587.93&  595.93& $70.46_{-2.51}^{+2.16}\rule{0mm}{1.2em}$ &$+0.019_{-0.255}^{+0.541}$ &
 $\alpha=0.647_{-0.64}^{+3.25}$,  \\
 31 $H(z)$  &  &  &  &  &  $B_s =0.826_{-0.111}^{+0.284} \rule{0mm}{1.2em}$   \\
  \hline
\end{tabular}
 \label{Estim}\end{table}

Below we concentrate on the more relevant summarized function $\chi^2_{tot}$. In
Fig.~\ref{F3} we present other two- and one-parameter distributions of  $\chi^2_{tot}$
and the likelihood functions for the $\Lambda$CDM model. In particular,  in the
top-right panel the contour plots for
$m^\chi_{tot}(\Omega_k,H_0)=\min\limits_{\Omega_m}\chi^2_{tot}$ are shown for the cases
$N_H=57$ and $N_H=31$ in the same notations. In these calculation we consider the
curvature fraction $\Omega_k$ as an independent parameter (together  with
$\Omega_m,H_0$), the fraction $\Omega_\Lambda$ is expressed via Eq.~(\ref{sumOm}):
$\Omega_\Lambda =1-\Omega_m-\Omega_k$.

The  two-parameter distributions (\ref{2paramL}) $m^\chi_{tot}(\Omega_m,\Omega_\Lambda)$
for  $N_H=57$ and 31 in the top-right panel of Figs.~\ref{F2}, \ref{F3} let us calculate
the one-parameter distributions $m^\chi_{tot}(\Omega_m)$, $m^\chi_{tot}(\Omega_\Lambda)$
and the likelihood functions (\ref{likeli}) ${\cal L}_{tot}(\Omega_m)$, ${\cal
L}_{tot}(\Omega_\Lambda)$  shown in the middle and  bottom panels of Fig.~\ref{F3}. The
functions ${\cal L}_{tot}(H_0)$  are deduces from the two-parameter distributions in the
$(\Omega_k,H_0)$ plane.

The best fitted values of $\min\chi^2_{tot}$ and the  model parameters $\Omega_m$,
$\Omega_\Lambda$, $\Omega_k$, $H_0$ for the $\Lambda$CDM  model are presented in
Table~\ref{Estim} for the cases $N_H=57$ and  $N_H=31$. The $1\sigma$ errors are
calculated from the correspondent  likelihood functions (\ref{likeli}) ${\cal
L}_{tot}(p_i)$. We should emphasize, that the  number $N_p$ of model parameters is
essential, when we comrade different models. So we also use the Akaike information
criterion \cite{Sharov16,ShiHL12}
  \begin{equation}
 AIC = \min\chi^2_{tot} +2N_p.
 \label{AIC} \end{equation}

 Here $N_p=3$ for the $\Lambda$CDM model.

The similar estimations for the $\Lambda$CDM model were made in many papers, in
particular, in Refs.~\cite{Planck15,Planck13,WMAP,Sharov16,ShiHL12,FarooqMR13,Jesus18}
for describing the Type Ia supernovae, $H(z)$, BAO and other data in various
combinations. One can observe the following effect (connected with the described above):
the estimations of $\Omega_m$, $\Omega_\Lambda$, $\Omega_k$ and $H_0$ in different papers
essentially depend on a chosen  $H(z)$ data set. For example, the authors of
Refs.~\cite{Jesus18} used the $\chi^2_H$ function with $N_H=41$ data points from both DA
and BAO methods and calculated $\Omega_m=0.237\pm0.051$,  $\Omega_\Lambda=0.66\pm0.20$.
However, when they excluded 3 data points \cite{Font-Ribera13,Delubac14,Busca12} with
$z\ge2.3$, they obtained the enhanced values for both parameters
$\Omega_m=0.40_{-0.14}^{+0.18}$, $\Omega_\Lambda=0.92_{-0.23}^{+0.34}$ (compare with our
results for $\chi^2_H$ in Fig.~\ref{F2}).

\begin{figure}[bh]
  \centerline{\includegraphics[scale=0.7,trim=6mm 5mm 5mm 3mm]{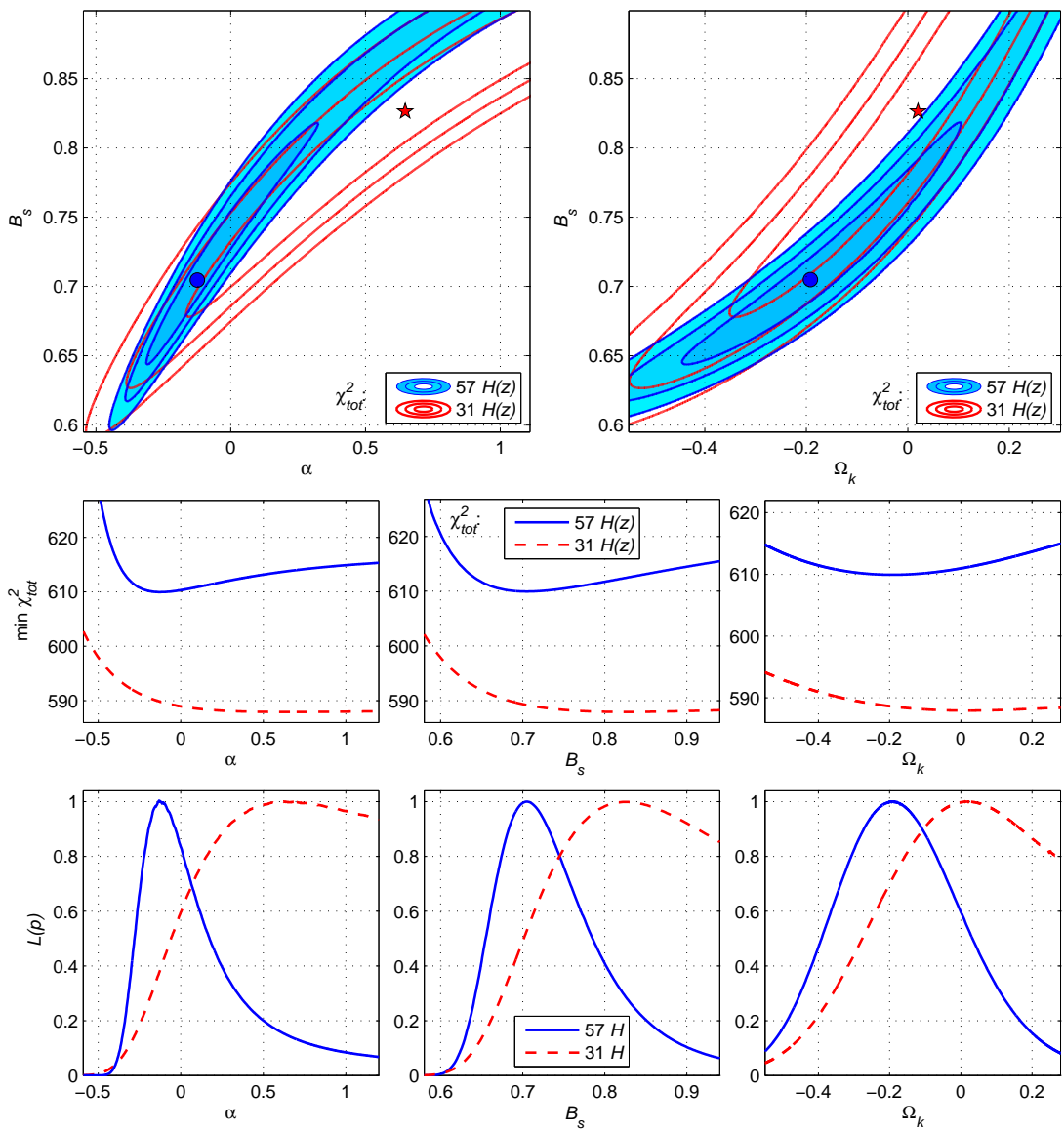}}
\caption{  The GCG model with $N_H=57$ (blue) and $N_H=31$ (red): two-parameter,
one-parameter distributions and likelihood functions for  $\chi^2_{tot}$. }
  \label{F4}
\end{figure}

If we compare our results for the $\Lambda$CDM model  with the latest Planck data \cite{Planck15}
($\Omega_m=0.308\pm 0.012$, $\Omega_\Lambda=0.692\pm 0.012$,  $\Omega_k=-0.005^{+0.016}_{-0.017}$,
$H_0=67.8\pm0.9$ km\,c${}^{-1}$Mpc${}^{-1}$), we will find some tension for $\Omega_\Lambda$,
$Omega_k$ in the case $N_H=31$ and for $H_0$ in both cases  because of too low estimation
of $H_0$ in Ref.~\cite{Planck15}.

\medskip

\begin{figure}[t]
  \centerline{\includegraphics[scale=0.71,trim=9mm 6mm 5mm 6mm]{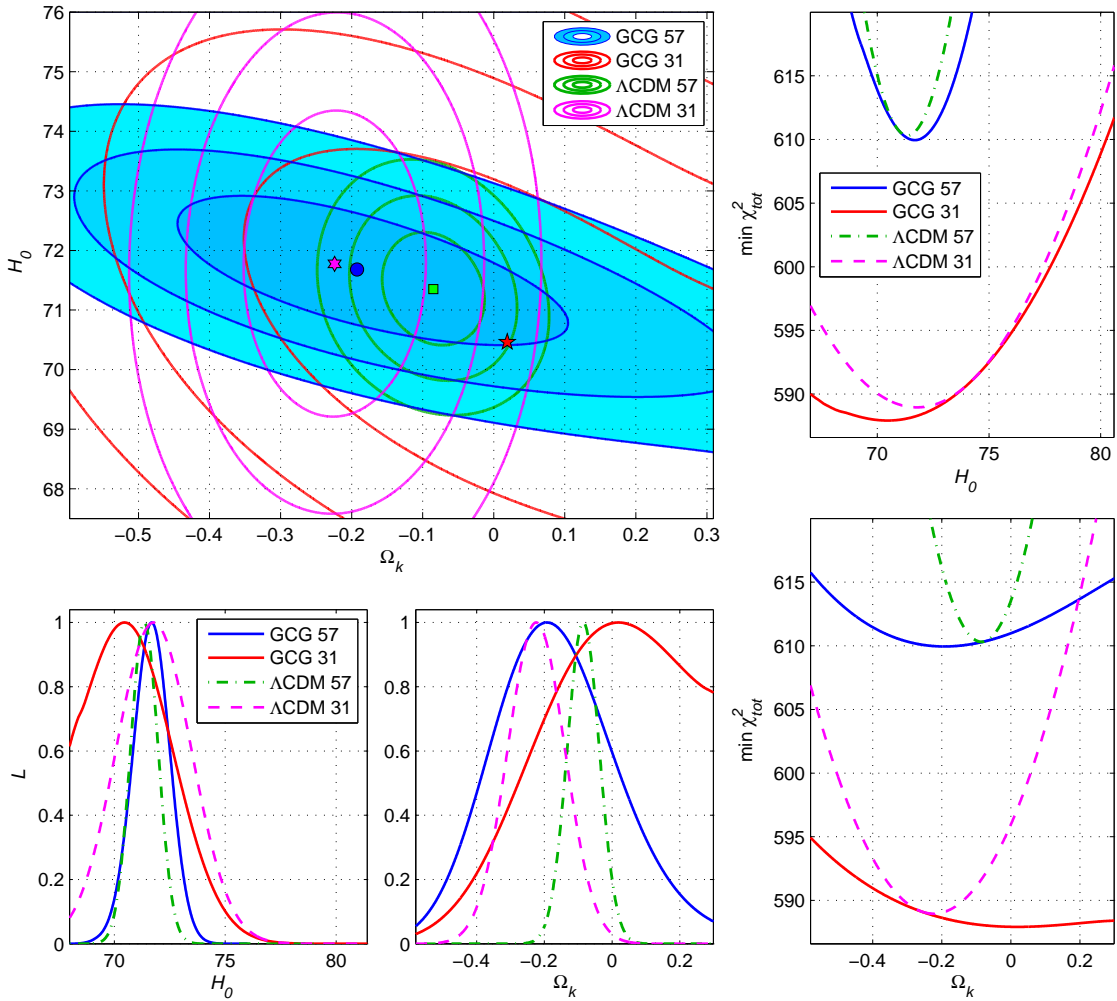}}
  \caption{\small Comparison of the two-parameter distributions $\min\chi^2_{tot}(\Omega_k,H_0)$
for the $\Lambda$CDM and GCG models in the plane $(\Omega_k,H_0)$ of their common
parameters for the cases with 57 and 31 $H(z)$ data points (the top-left panel). The
corresponding one-parameter  distributions are in other panels. Notations correspond to
the previous figures. }
  \label{F5}
\end{figure}

The influence of a chosen  $H(z)$ data set takes place not only for the $\Lambda$CDM
model. One can see in Fig.~\ref{F4} and in Table~\ref{Estim}, that for the GCG model
this influence is even more strong. In the top panels we demonstrate the contour plots
for two-parameter distributions (\ref{2param}) of $\chi^2_{tot}$ in the $(\alpha,B_s)$
and $(\Omega_k,B_s)$ planes for the cases $N_H=57$ (blue filled contours) and $N_H=31$
(red contours). In particular, the two-parameter distributions (\ref{2param}) in the
top-left panel are
 $$
 m^\chi_{tot}(\alpha,B_s)=\min\limits_{\Omega_k,H_0}\chi^2_{tot}(\alpha,B_s,\Omega_k,H_0).
 $$
The circles and stars show the points of minima for  $\chi^2_{tot}$. The similar
two-parameter contour plots for the GCG model in the $(\Omega_k,H_0)$ plane are drawn in
Fig.~\ref{F5}.

The one-parameter distributions $m^\chi_{tot}(\alpha)$, $m^\chi_{tot}(B_s)$,
$m^\chi_{tot}(\Omega_k)$  and the corresponding likelihood functions (\ref{likeli})
${\cal L}_{tot}(p_i)$ are  shown in the middle and bottom panels of Fig.~\ref{F4}.

Fig.~\ref{F4} and Table~\ref{Estim} demonstrate, that for the GCG model the best fitted
values of $\alpha$, $B_s$, $\Omega_k$ strongly depend on a Hubble parameter data:
$N_H=57$ (all data points) or $N_H=31$ (only from DA method). In particular, the best
fitted values $\alpha\simeq-0.124$,  $\Omega_k\simeq-0.192$ for $N_H=57$ change their
signs and become  $\alpha\simeq+0.647$,  $\Omega_k\simeq+0.019$, if $N_H=31$.

In  Fig.~\ref{F5} we compare the $\Lambda$CDM and GCG models in the plane
$(\Omega_k,H_0)$ of their common parameters. For both models we draw the one-parameter
distributions  $m^\chi_{tot}(\Omega_k)$, $m^\chi_{tot}(H_0)$ (they help us to compare
the best results $\min\chi_{tot}$ for these models) and the likelihood functions
 ${\cal L}_{tot}(\Omega_k)$, ${\cal L}_{tot}(H_0)$.

In the top-left panel of Fig.~\ref{F5} the filled contours describe the GCG model with
$N_H=57$, other contours differ in their color. The points of minima are marked here as
the circle (GCG, $N_H=57$), the pentagram (GCG, $N_H=31$), the square ($\Lambda$CDM,
$N_H=57$) and the hexagrams ($\Lambda$CDM, $N_H=31$) of the corresponding color.

Fig.~\ref{F5} is useful, when we want to compare predictions the $\Lambda$CDM and GCG
models in the considered cases  $N_H=57$ and $N_H=31$. The plots  ${\cal
L}_{tot}(\Omega_k)$ and ${\cal L}_{tot}(H_0)$ show differences of the best fitted
values, the plots  $m^\chi_{tot}(\Omega_k)$ and $m^\chi_{tot}(H_0)$ describe
effectiveness of these models. Mere detailed information is tabulated in
Table~\ref{Estim}.

\section{Conclusion}

In this paper we describe the observational data for Type Ia supernovae \cite{SNTable},
BAO (Table~\ref{TBAO}) and two data sets of the Hubble parameter data $H(z)$ (all
$N_H=57$ data points from Table~\ref{TH} and only 31  data points from differential
ages) with the $\Lambda$CDM model and the model with generalized Chaplygin gas (GCG).

The results are demonstrated in Table~\ref{Estim}: for all models and variants of $N_H$
we calculated the minimal values of the function $\chi^2_{tot}$ (\ref{chitot}),  the
results  of Akaike information criterion (\ref{AIC}) and the best fitted values of model parameters
with $1\sigma$ errors. For the GCG model we achieve the best minimal values of $\min\chi^2_{tot}$,
however the  Akaike criterion gives advantage to the $\Lambda$CDM model, because it has the
small number $N_p=3$ of model parameters (degrees of freedom) in comparison with with $N_p=4$
for GCG.

But the most striking result of our calculations for both models is the large difference between
the best fitted values of model parameters in the cases with $N_H=57$ $H(z)$ data points
from Table~\ref{TH} and  $N_H=31$ data points, obtained with DA method
(the left hand side of  Table~\ref{TH}). For the case $N_H=57$ these results are close to the
estimations for these models in Ref.~\cite{Sharov16}, because in that paper we used $H(z)$ data points from
both DA and BAO methods (though there were $N_H=38$ points).

This essential divergence between the predictions of the variants with all $N_H=57$ and
$N_H=31$ DA data points is seen visually in Fig.~\ref{F1}. It may be explained and
connected with 4 $H(z)$ data points \cite{Font-Ribera13,Delubac14,Busca12,Bautista17}
with high redshifts $z\ge2.3$. These data points, obtained with BAO method (see the
right hand side of  Table~\ref{TH}) have small errors $\sigma_H$ and strongly influence
on a model predictions, when we take these points into account (in the case $N_H=57$).
Otherwise, when we include only $N_H=31$ DA data points, this effect disappears.

\end{document}